# Particle Geometry Space: An integrated characterization of particle shape, surface area, volume, specific surface, and size distribution

Priya Tripathi, and Seung Jae Lee*

Department of Civil & Environmental Engineering, Florida International University, Miami, FL, USA

## Abstract

Particle *size* and *shape* are the key 3D particle geometry parameters that govern the complex behavior of granular materials. The effect of particle size and shape has often been examined in isolation, typically through separate analyses of particle size distribution (PSD) and shape distribution, leading to an unaddressed knowledge gap. Beyond size and shape, 3D particle geometry also includes attributes such as *surface area* and *volume*, which together defines the *surface-area-to-volume ratio*, commonly known as the *specific surface*. To comprehensively understand the influence of particle geometry on the behavior of granular materials, it is important to integrate these parameters, ideally into a single analytical framework. To this end, this paper presents a new approach, *particle geometry space* (*PGS*), formulated based on the principle that the key 3D particle geometry attributes – *volume*, *surface area*, and *shape* – can be uniformly expressed as a function of *specific surface*. The PGS not only encompasses all 3D particle geometry attributes but also extends its scope by integrating the conventional PSD concept. This innovation enables engineers and researchers who are already familiar with PSD to perform a more systematic characterization of 3D particle geometries. The paper (i) discusses the limitations of existing methods for characterizing 3D particle geometry, (ii) offers an overview of the PGS, (iii) proposes a method for integrating PSD into the PGS, and (iv) demonstrates its application with a set of 3D mineral particle geometry data.

Keywords: 3D particle geometry; Particle geometry space (PGS); Shape; Surface area; Volume; Specific surface; Particle size distribution (PSD)

*Corresponding author: sjlee@fiu.edu

# 1 INTRODUCTION

In geotechnical engineering, the characterization of granular materials has traditionally relied on particle size distribution (PSD). Beyond that, 3D particle shape is an equally important particle-scale property that significantly influences the macroscopic behavior of granular materials. For example, track ballast – a granular material essential for railroad infrastructure – performs critical functions such as facilitating drainage around the tracks and distributing the load from train traffic. The geometry of the ballast particles including size and shape is a key microscopic parameter governing these functions, influencing various macroscopic aspects of the track performance [1–3]. Furthermore, the ballast particle geometries undergo changes over time due to fouling, caused by particle abrasion and breakage [4,5]. The gradual particle geometry changes lead to a decline in track performance, emphasizing the importance of systematic characterization of 3D particle geometries to enable robust track ballast monitoring and maintenance. Therefore, efforts to characterize particle geometries – both size and shape – have been emphasized within the research community. The most common practice involves presenting the distributions of both size and shape, as illustrated in Figure 1, where shape is characterized by Wadell's true sphericity [6], denoted as $S$ in the figure. This presentation has been broadly adopted in the fields related to granular materials [7–14].

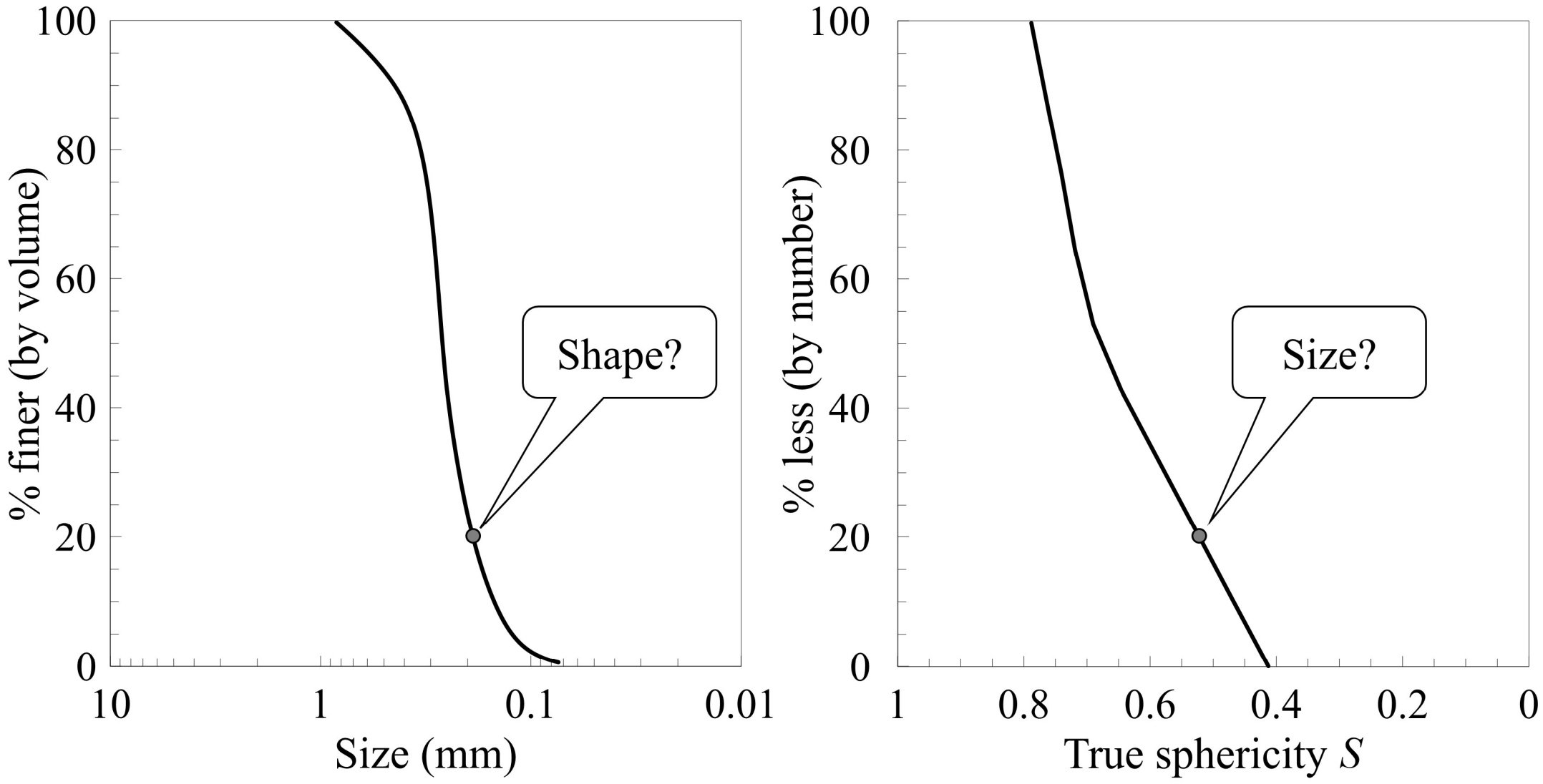

Figure 1. Separate presentation of particle size and shape distributions.

Although commonly adopted, presenting particle shape and size separately, as in Figure 1, limits comprehensive geometry characterization; the information about particle shape and size is not cross-referenced with each other. For example, the shape corresponding to size $D_{20}$ (highlighted in Figure 1) cannot be determined despite the shape distribution available at hand. Likewise, it is not possible to determine the size associated with a specific shape of interest.

Addressing this limitation is important for the accurate characterization of granular materials. As an example, Figure 2a illustrates three sets of particles, each sorted by their size denoted as $D$. All sets contain particles ranging in size from 10 to 1. Therefore, these sets have identical size distributions, as shown in Figure 2b. The shapes are quantified using true sphericity ($S$), a measure that evaluates how closely a particle's surface area ($A$) compares to that of a sphere ($A_s$) with the same volume, as defined in Equation (1). $S$ ranges from 0 to 1, with $S = 1$ representing a perfect sphere. The three sets in the figure contain identical shapes – three tetrahedrons, three cubes, and four spheres – resulting in the identical shape distributions as shown in Figure 2c. The data points correspond to a composition of 30% tetrahedrons ($S = 0.67$), 30% cubes ($S = 0.81$), and 40% spheres ($S = 1$).

$$S = A_s / A \tag{1}$$

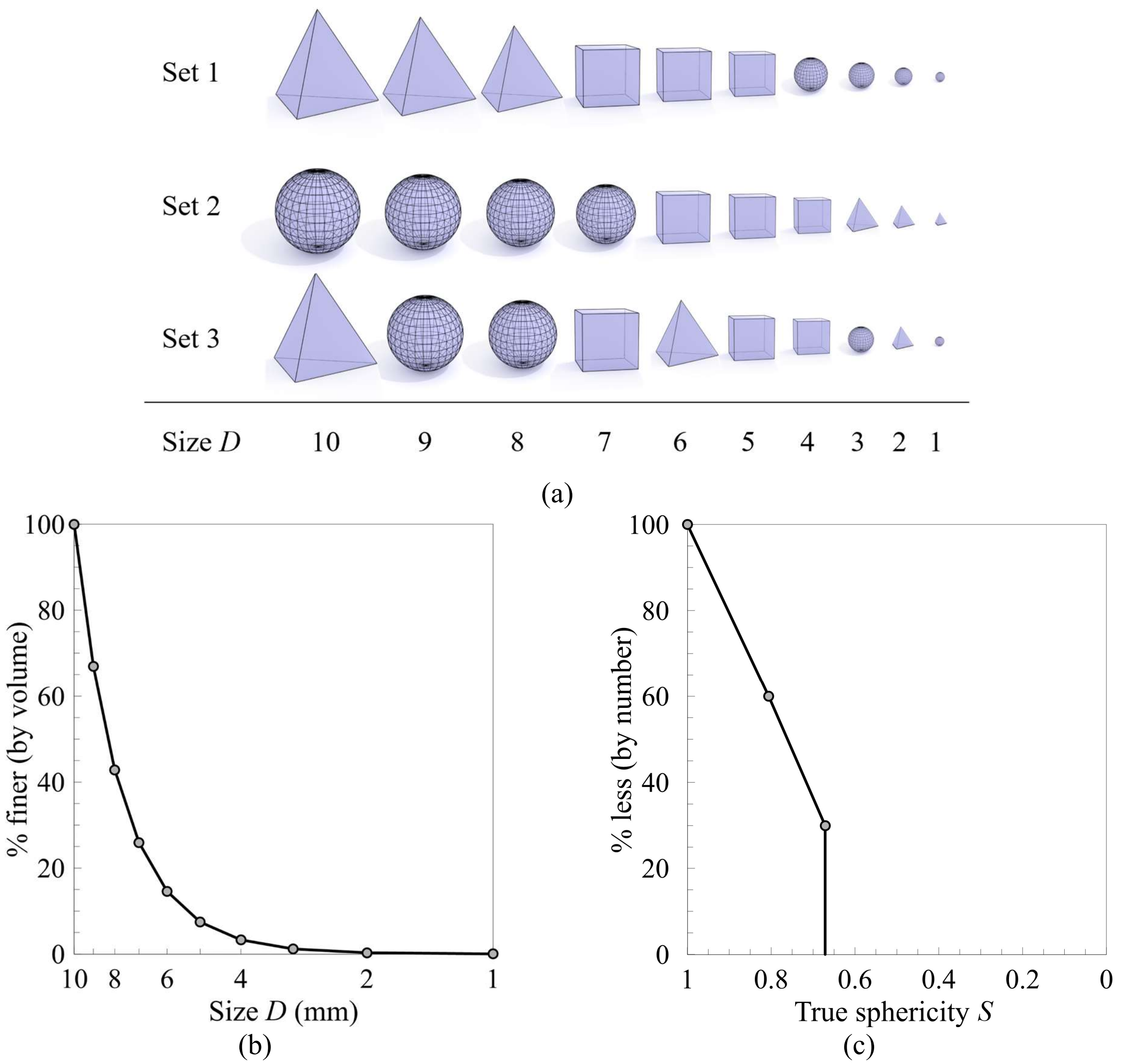

Figure 2. Limitation of characterizing particle sizes and shapes in isolation: (a) Three different particle sets; (b) Identical particle size distributions; and (c) Identical particle shape distributions.

Despite having identical size and shape distributions as shown in Figure 2b and c, these sets clearly differ from one another. In the set 1, the spheres are the smallest, while the tetrahedrons are the largest. In contrast, the set 2 exhibits the reverse pattern. On the other hand, the particles in the set 3 do not have a specific pattern. Presenting the size and shape distributions as depicted in Figure 2b and c fails to distinguish between these sets. This example highlights a critical limitation of characterizing particle sizes and shapes in isolation, emphasizing the necessity to correlate these geometric attributes.

An alternative approach to addressing this limitation was presented by Altuhafi and Coop [15]. Their study explicitly presents a relationship between shape and size alongside PSD (see Fig. 15). Using their approach, Figure 1, which presents shape and size separately, can be transformed into Figure 3. This plot shows that the smallest particle has a true sphericity (*S*) value of around 0.4, while the largest particle measures about 0.8. This presentation reveals a trend in which larger particles tend to be more spherical than smaller ones. Therefore, with this approach, the relationship between the particle shape and size can be better identified.

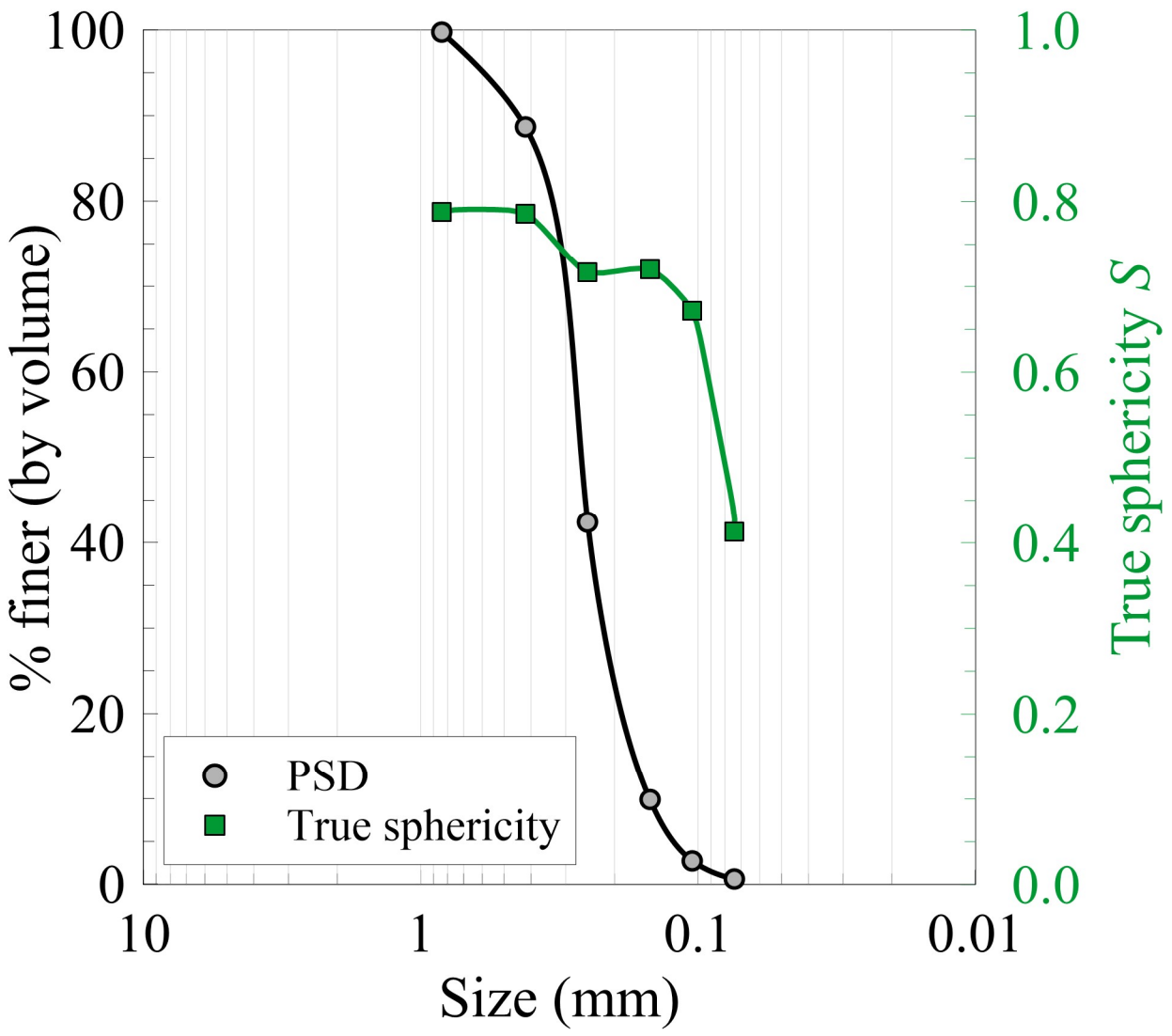

Figure 3. Particle size distribution (PSD) presented in relation to shape.

While the method of presenting PSD and its relationship to particle shape as in Altuhafi and Coop [15] offers clear benefits, it prompts a compelling question for further exploration: Is it feasible to integrate all 3D particle geometry attributes – including shape, size, volume, surface area, and specific surface – be unified within a single plot space to effectively illustrate their interrelationships? Such an approach could provide a more comprehensive visualization of how these attributes interrelate, potentially offering deeper insights into the effects of 3D particle geometry on the behavior of granular materials.

This paper introduces a significant breakthrough with a new approach that seamlessly integrates all 3D particle geometry attributes – shape, surface area, volume, and specific surface – along with PSD, all within a single framework. In pursuit of the goal, the study presents the *particle geometry space* (*PGS*), initially conceived in the authors' previous work [16,17], where it was referred to as "logarithmic $A/V$ and $V$ space." This paper expands the concept into a formal theory by elucidating 3D particle geometry as a function of the specific surface. Section 2 introduces the theory behind particle geometry space and discusses its integration with PSD. Section 3 demonstrates the approach using a set of 3D mineral particle geometry data.

# 2 METHODOLOGY

## 2.1 Particle geometry space (PGS) – Single particle geometry

The primary attributes of a 3D particle geometry include *size* ($D$), *volume* ($V$), *surface area* ($A$), and *shape* ($\beta$), along with derived attributes, the *surface area-to-volume ratio* ($A/V$). Size and volume are intrinsically related, with the size ($D$) expressed as the diameter of a sphere with the same volume ($V$) as the particle, as shown in Equation (2).

Other geometry attributes can be uniformly presented by $\mathcal{g} = (A/V)^k \times V$, as shown in Equations (3) to (5), with variations occurring only in the exponent ($k$) applied to $A/V$. The $A/V$ ratio is also referred to as *specific surface* [18]. The parameter $\beta$ serves as a *shape* index and is related to Wadell's true sphericity ($S$), a classical shape index widely used since its inception in the 1930s [19]. The parameter $\beta$ is essentially the inverse of $S$, as shown in Equation (6).

$$D = (6V / \pi)^{1/3} \tag{2}$$

$$V = (A/V)^0 \times V \tag{3}$$

$$A = (A/V)^1 \times V \tag{4}$$

$$\beta = (A/V)^3 \times V \tag{5}$$

$$(\beta / 36\pi)^{1/3} = S^{-1} \tag{6}$$

The derivation of Equation (6) is presented below:

| | | |
|---|---|---|
| $(\beta / 36\pi)^{1/3}$ | $= ((A^3 / V^2) / 36\pi)^{1/3}$ | Since $\beta = A^3 / V^2$, as shown in Equation (5), |
| | $= A / (V_s^2 \times 36\pi)^{1/3}$ | Considering a sphere having the same volume as the particle, i.e., $V = V_s$, where $V_s$ represents the sphere's volume, |
| | $= A / A_s$ | where $A_s = (V_s^2 \times 36\pi)^{1/3}$, based on the theoretical surface area to volume relationship of a sphere, |
| | $= S^{-1}$ | according to Equation (1). |

Due to the inverse relationship between $\beta$ and $S$, as a particle deviates further from a spherical shape, $S$ decreases while $\beta$ increases. For example, a sphere ($S = 1$) has $\beta = 36\pi$ (~113), which is the lowest possible $\beta$ value. In contrast, a cube ($S = 0.81$) has $\beta = 216$, and a tetrahedron ($S = 0.67$) has $\beta = 374$. Figure 4 provides additional examples of particle shapes evaluated using $S$ and the new shape index ($\beta$) for various well-known polyhedrons and selected mineral particles. Since $\beta = 36\pi$ for a sphere, dividing $\beta$ by $36\pi$ represents a particle's shape relative to a sphere.

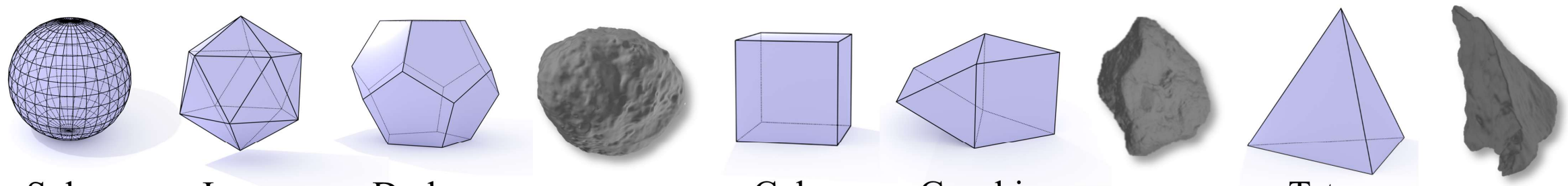

| | Sphere | Icosa-hedron | Dodeca-hedron | | Cube | Gyrobi-fastigium | | Tetra-hedron | |
|---|---|---|---|---|---|---|---|---|---|
| $S$: | 1.00 | 0.94 | 0.91 | 0.87 | 0.81 | 0.77 | 0.72 | 0.67 | 0.56 |
| $\beta$: | $36\pi$ | 136 | 150 | 172 | 216 | 251 | 303 | 374 | 644 |
| $\beta / 36\pi$: | 1.00 | 1.21 | 1.33 | 1.52 | 1.91 | 2.22 | 2.68 | 3.31 | 5.69 |

Figure 4. Examples of particle shape evaluated with true sphericity ($S$) and the shape index $\beta$.

According to Equations (3) to (5), the geometric attributes of particles primarily depend on the specific surface ($A/V$). This finding highlights that specific surface is a key geometric parameter influencing macroscopic granular material behaviors such as modulus, strength, permeability, and transport processes, as observed in numerous studies [20–23]. For example, fine-grained soils retain more moisture than coarse-grained soils, making their behaviors more sensitive to moisture. Specific surface is inversely proportional to particle size (its unit is reciprocal length, e.g., $mm^{-1}$), so a mass of fine particles has a much greater specific surface than an equal mass of coarse particles. Moreover, particle shape is inherently linked to specific surface (according to Equation (5)). As a result, shape-related effects are significant for particles of all sizes, even for fine particles. For instance, among common clay minerals, kaolinite has the thickest particle shape, illite has intermediate particle thickness, and montmorillonite has the thinnest. Consequently, kaolinite exhibits the smallest specific surface, whereas montmorillonite has the largest [18]. These shape differences are linked to specific surface, which directly influences how each mineral interacts with water, leading to distinct macroscopic soil behaviors.

Log-transforming the four equations from (3) to (5) results in linear relationships ($y = ax + b$), as shown in Equations (7) to (9), in a log-log space, where $x$-axis denotes $\log(A/V)$, and $y$-axis presents $\log(V)$. This log-transformed space defines the *particle geometry space* (*PGS*), as shown in Figure 5. An auxiliary $y$-axis for $\log(D)$ can be introduced in addition to $\log(V)$, following the relationship in Equation (2). The specific placement of the $\log(D)$ axis along the horizontal direction is arbitrary and does not affect the interpretation of the results. Lines with *slopes* 0, -1, and -3 intersect at $A/V = 1$. These *intercepts* at $A/V = 1$ correspond to $V$, $A$, and $\beta$. Therefore, PGS effectively integrates all 3D particle geometry attributes within a unified framework. Geometrically, $\log(\beta/A)$ is twice $\log(A/V)$ in the PGS because the two triangles in Figure 5 share the same base length, but one has a slope three times steeper than the other. This leads to Equation (10), which can be reformulated as Equation (5).

$$\log(V) = 0 \times \log(A/V) + \log(V) \tag{7}$$

$$\log(V) = -1 \times \log(A/V) + \log(A) \tag{8}$$

$$\log(V) = -3 \times \log(A/V) + \log(\beta) \tag{9}$$

$$\log(\beta/A) = 2 \times \log(A/V) \tag{10}$$

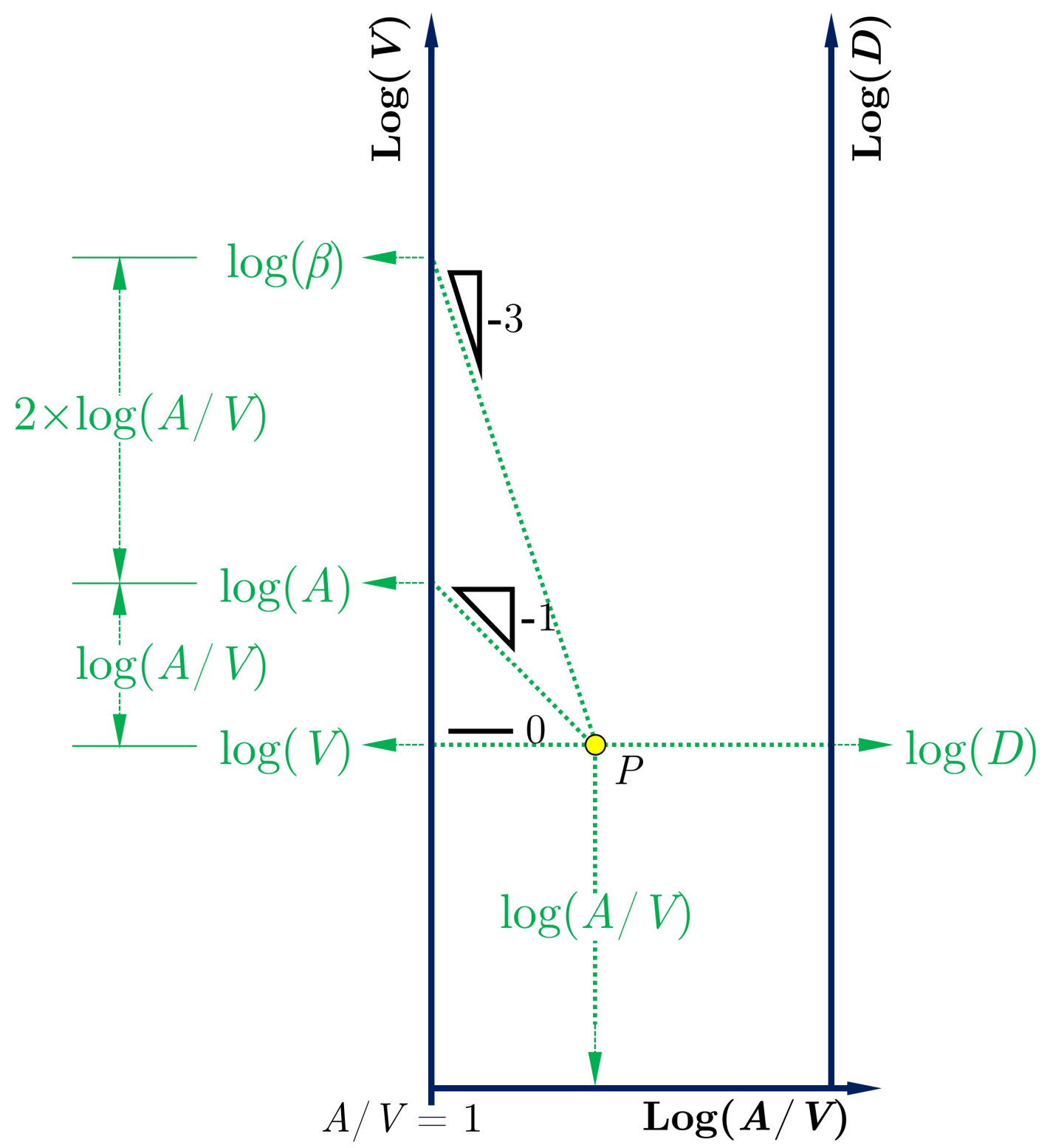

Figure 5. Particle geometry space (PGS): All geometry attributes ($V$, $A$, $A/V$, $D$, and $\beta$) can be holistically presented. The yellow dot represents a particle data point. For simplicity, hereafter, all symbols within the PGS presented hereafter will be denoted without the log( ) notation.

The following demonstrates its applicability to idealized and real-world particles. The first example is shown in Figure 6a. For an illustrative purpose, a regular tetrahedron (with all four faces as equilateral triangles) is considered, where the theoretical relationships between the geometric attributes are well-defined.

a) Theoretical values: The volume of the regular tetrahedron is assumed to be $V$ = 5 mm³, with a corresponding surface area of $A$ = 21.1 mm², resulting in $A/V$ = 4.21 mm⁻¹. Using Equations (2) and (5), the particle size is $D$ = 2.12 mm, and the shape index is determined to be $\beta$ = 374.
b) Placement of data point into the Particle Geometry Space (PGS): The data corresponding to the particle is plotted within the PGS using the $A/V$ and $V$ values.
c) Determination of geometric attributes within the PGS: From the plotted data point, intercepts at $A/V$ = 1 are determined along lines with slopes of -1 and -3, which correspond to $A$ and $\beta$, respectively. The intercepts confirm that $A$ = 21.1 mm² and $\beta$ = 374. The horizontal intersection with the log($D$) axis corresponds to the particle size $D$ = 2.12 mm.

The second example (Figure 6b) applies the method to a mineral particle, demonstrating its effectiveness in analyzing naturally occurring, irregular geometries.

a) 3D scanning of a mineral particle: A 3D scan is performed on a selected mineral particle using a Polyga C504 structured light 3D scanner [24] to capture its geometric attributes. The scanned particle, identified as VG_b_01, belongs to the ‘Virginia Granite - Group B’ dataset available in the NSF DesignSafe-CI repository [25]. The measured parameters include a surface area of $A$ =

518.6 mm² and a volume of $V$ = 646 mm³, yielding $A/V$ = 0.8 mm⁻¹. The shape index is determined to be $\beta$ = 334, and the particle size is $D$ = 10.73 mm.

b) Data placement and determination of geometric attributes in the PGS: The scanned particle's geometric data is mapped within the PGS. The intercepts at $A/V$ = 1 confirm that $A$ and $\beta$ values obtained above. The horizontal intersection with the log($D$) axis corresponds to $D$ = 10.73 mm.

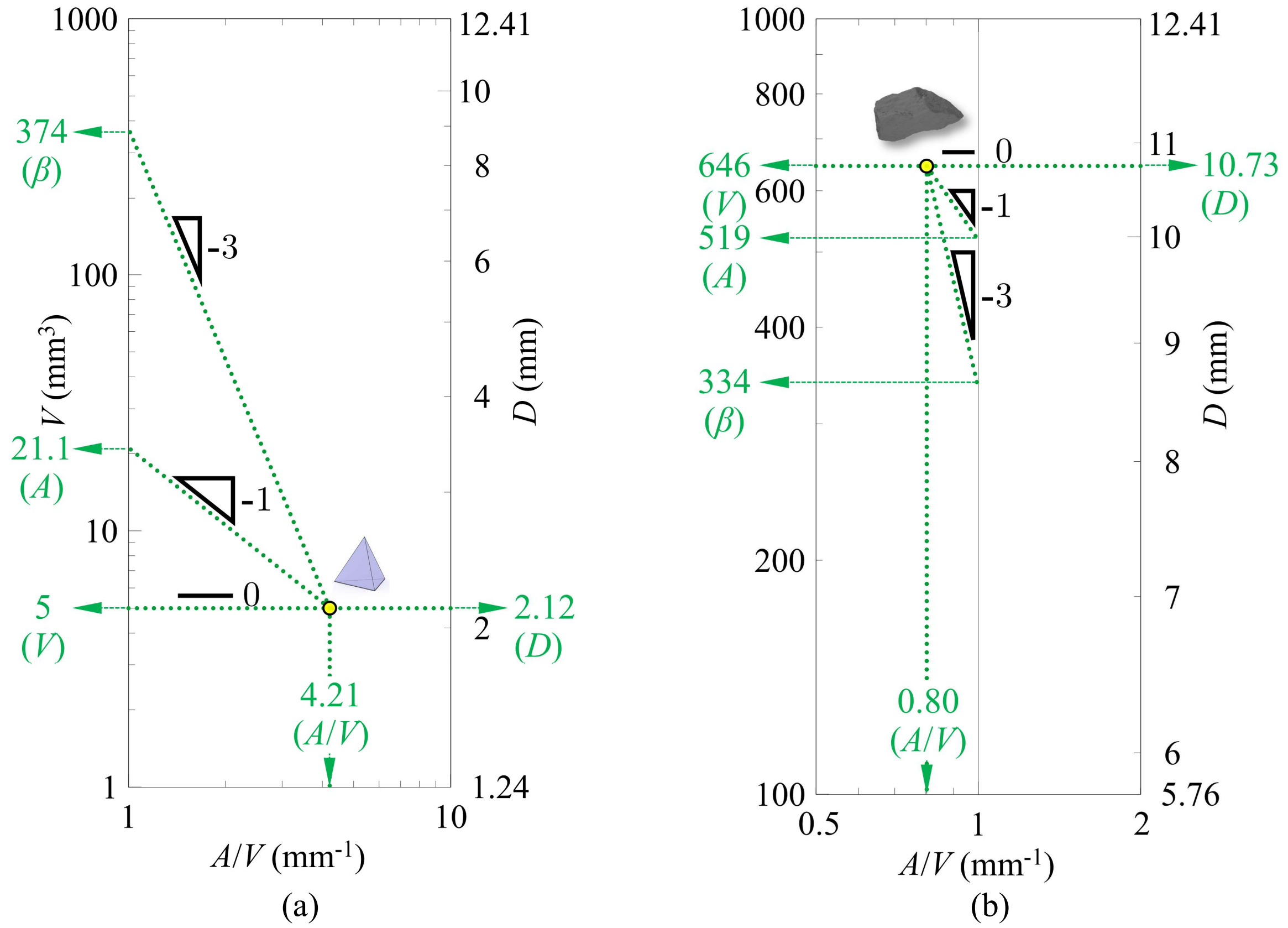

Figure 6. Demonstration of PGS: (a) A tetrahedron; and (b) A mineral particle.

A perfect sphere always aligns with the *sphere line*, where $\beta = 36\pi$, as shown in Figure 7. Any data point on this sphere line represents a sphere. Graphically, being farther from the sphere line indicates a larger deviation from a spherical shape. Lines for other specific shapes, such as the *cube line* and *tetrahedron line*, can also be defined with their corresponding $\beta$ values of 216 and 374, respectively. Since no shape can be rounder than a sphere, no data points can appear in the gray-shaded area labeled as *Iz* (impossible zone) below the sphere line.

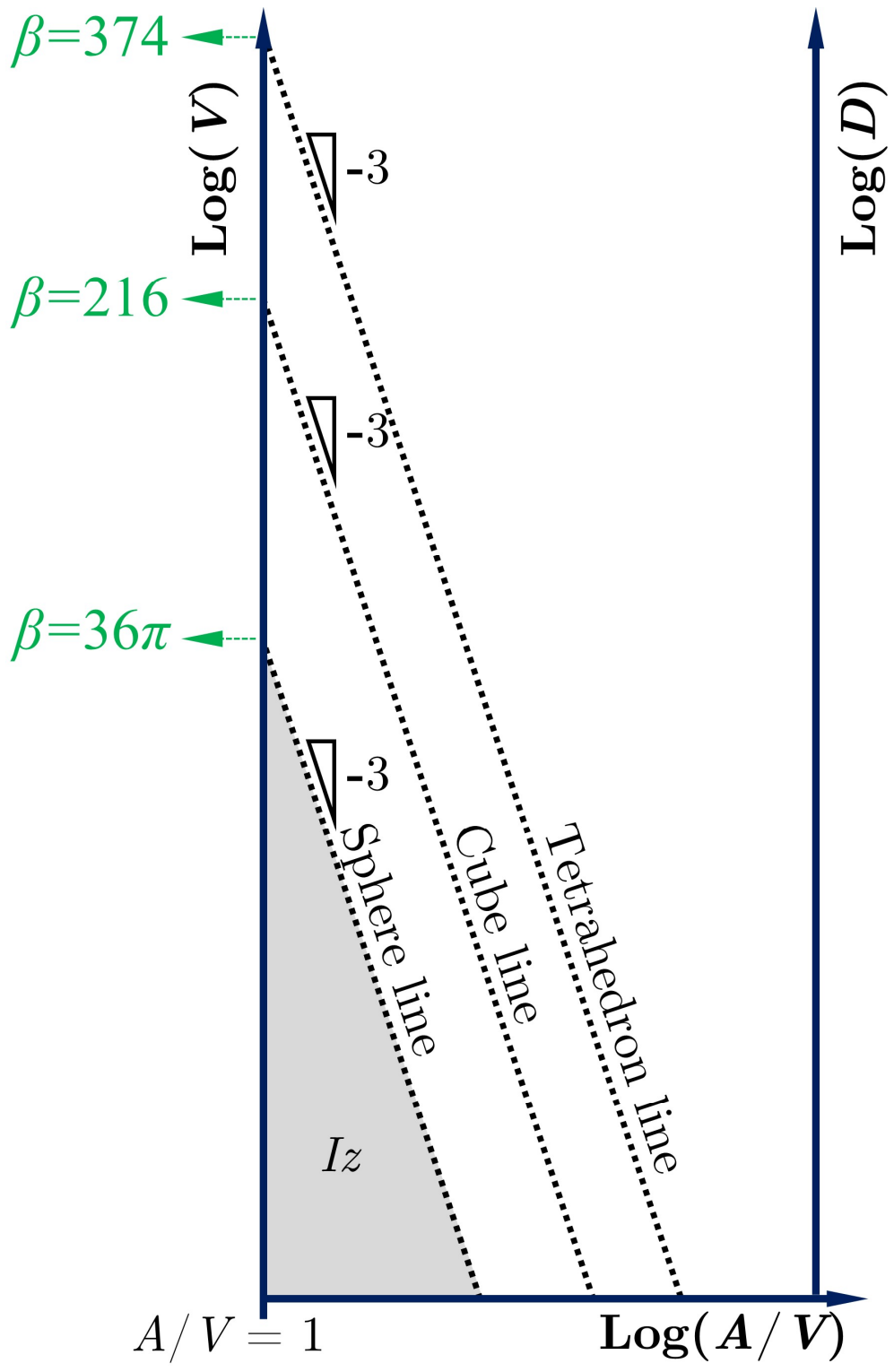

Figure 7. The lower bound in PGS, defined by the sphere line; The lowest possible $\beta$ is $36\pi$, representing a sphere, and no data can exist in the *Iz* area.

Figure 8 presents an example featuring three particle data points, P1, P2, and P3, to demonstrate how the PGS can be utilized to compare geometric attributes of multiple particles. The corresponding shape indices are denoted as $\beta_{\mathrm{P1}}$, $\beta_{\mathrm{P2}}$, and $\beta_{\mathrm{P3}}$, respectively. Since P3 lies on the sphere line, it presents a perfect sphere with $\beta_{\mathrm{P3}} = 36\pi$. Given that P1 is positioned farther from the sphere line than P2, the shape indices follow the relationship $\beta_{\mathrm{P3}} < \beta_{\mathrm{P2}} < \beta_{\mathrm{P1}}$, indicating that P2 is more spherical than P1.

Other geometric attributes can be analyzed in a similar fashion. As the data points align along a line with a slope of -1, the surface areas are equal for all three particles, such that $A_{\mathrm{P3}} = A_{\mathrm{P2}} = A_{\mathrm{P1}}$. However, along the y-axis, P3 is positioned higher than P2, followed by P1, leading to the volume relationship $V_{\mathrm{P3}} > V_{\mathrm{P2}} > V_{\mathrm{P1}}$. Likewise, the particle sizes follow the same trend, with $D_{\mathrm{P3}} > D_{\mathrm{P2}} > D_{\mathrm{P1}}$. Along the *x*-axis, a trend of increasing specific surface is observed, following the order $(A/V)_{\mathrm{P3}} < (A/V)_{\mathrm{P2}} < (A/V)_{\mathrm{P1}}$, indicating that P1 has the highest relative surface area for its volume, while P3, being a sphere, has the lowest.

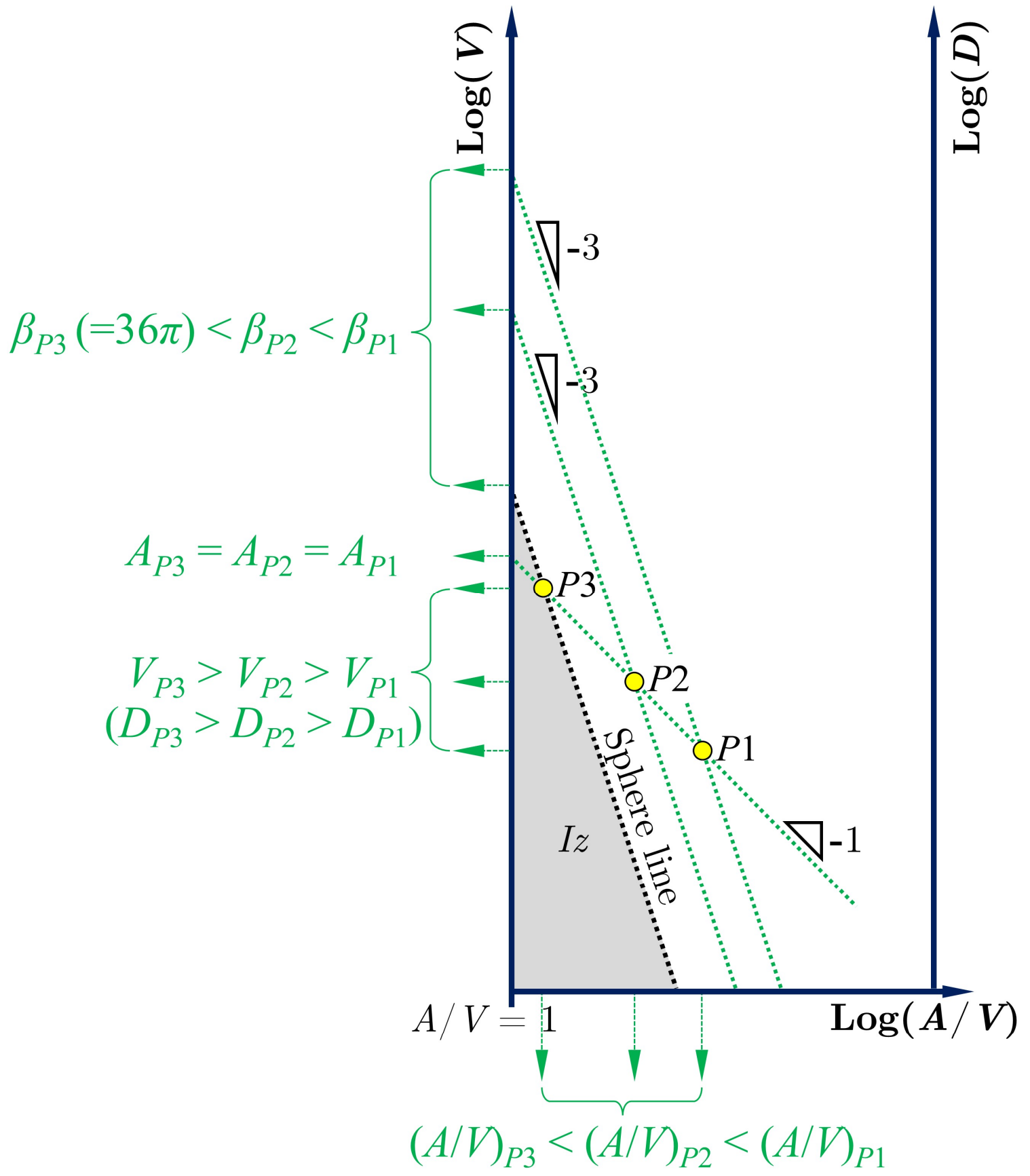

Figure 8. Comparison of geometry attributes within PGS.

## 2.2 Particle geometry space (PGS) – Geometries of particle population

When analyzing the geometries of particle population, which consists of many particles, statistical measures can be used to capture both (i) the average shape and (ii) the relationship between shape and size, such as how shape varies with size: (i) The average shape index $\beta^*$ can be analytically obtained from the $\beta$ values (of individual particles) as in Equation (11); (ii) The relationship between shape and size can be categorized into three cases: (a) Case 1: There is no specific relationship between shape and size; (b) Case 2: Larger particles tend to be more spherical than smaller ones; and (c) Case 3: Smaller particles tend to be more spherical than larger ones. When these cases are presented in the PGS, they appear as three distinct slopes.

$$\log(\beta^*) = \overline{\log(\beta)} \tag{11}$$

where $\overline{\log(\beta)}$ indicates the arithmetic mean of log ($\beta$).

Figure 9 illustrates how both (i) the average shape and (ii) the shape-size relationship are interpreted in the particle geometry space: (i) The $\beta^*$ averages the $\beta$ values. Therefore, similar to $\beta$, it is depicted graphically as the intercept at $A/V = 1$ on a line with a slope of -3; (ii) The three shape-size relationships are characterized by the distinct slopes, $\alpha$, of the *power regression line* of the geometry data, where $\alpha$ = -3, $\alpha$ > -3, and $\alpha$ < -3, respectively. For instance, consider the particle set illustrated in

Figure 9b. Here, $\beta_2$ corresponds to the largest particle's shape, while $\beta_1$ represents the smallest particle's shape. Notably, $\beta_2$ is smaller and closer to the sphere line than $\beta_1$, indicating the larger particle is more spherical than the smaller one. This results in a slope gentler than -3 ($\alpha > -3$). Despite identical average shapes across all cases (same $\beta^*$) in this specific example, variations in the slopes indicate distinct particle sets. Note that the slopes for all three cases are negative because $A/V$ and $V$ are inversely related. For example, for a given particle shape, when the size decreases, the $A/V$ ratio increases, resulting in a negative slope. Further discussion of the concept and details on prior experimental studies can be found in the referenced articles [5,16,25–27].

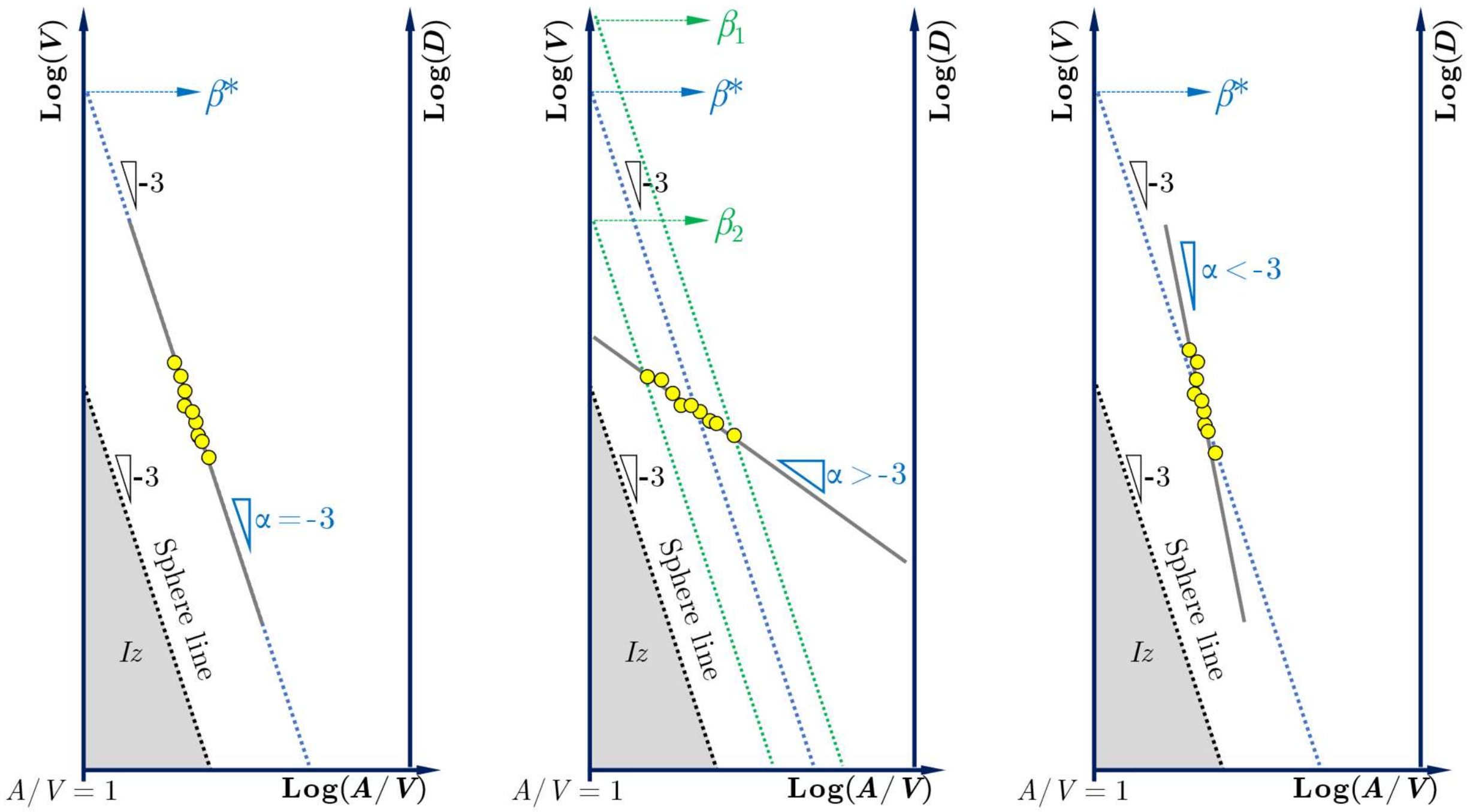

(a) Case 1 ($\alpha$ = -3) (b) Case 2 ($\alpha$ > -3) (c) Case 3 ($\alpha$ < -3)

Figure 9. Three cases of different shape-size relationships appear as distinct slopes, denoted by $\alpha$.

## 2.3 Integration of particle size distribution (PSD) into particle geometry space (PGS)

PSD can be integrated into the PGS by aligning it with the identified power regression line. First, the power regression line is derived from particle geometry data points plotted within the PGS, as illustrated in Figure 10a. Both PSD and PGS share a log($D$) axis, which are the unifying parameter needed for the integration. PSD is bounded by $D_{100}$ and $D_0$ (Figure 10b). These boundaries can be used as a reference for PSD scaling to integrate it into PGS, as demonstrated in Figure 10c. It is worth noting that using $D_{100}$ and $D_0$ is not necessary, and other size information can also be utilized. Once integrated, this approach enables the determination of other 3D particle geometry attributes for any size within the PSD. Figure 11 presents an example demonstrating how the attributes corresponding to the particle size $D_{60}$ are determined. The shape index $\beta_{D_{60}}$, which represents the shape associated with $D_{60}$, can be found from a line with slope of -3. Similarly, the surface area ($A_{D_{60}}$), the volume ($V_{D_{60}}$), the specific surface ($(A/V)_{D_{60}}$) corresponding to the specific particle size can be determined. In summary, all particle geometry attributes – including volume ($V$), surface area ($A$), specific surface ($A/V$), shape ($\beta$), size ($D$), as well as PSD – can be comprehensively presented within a single space.

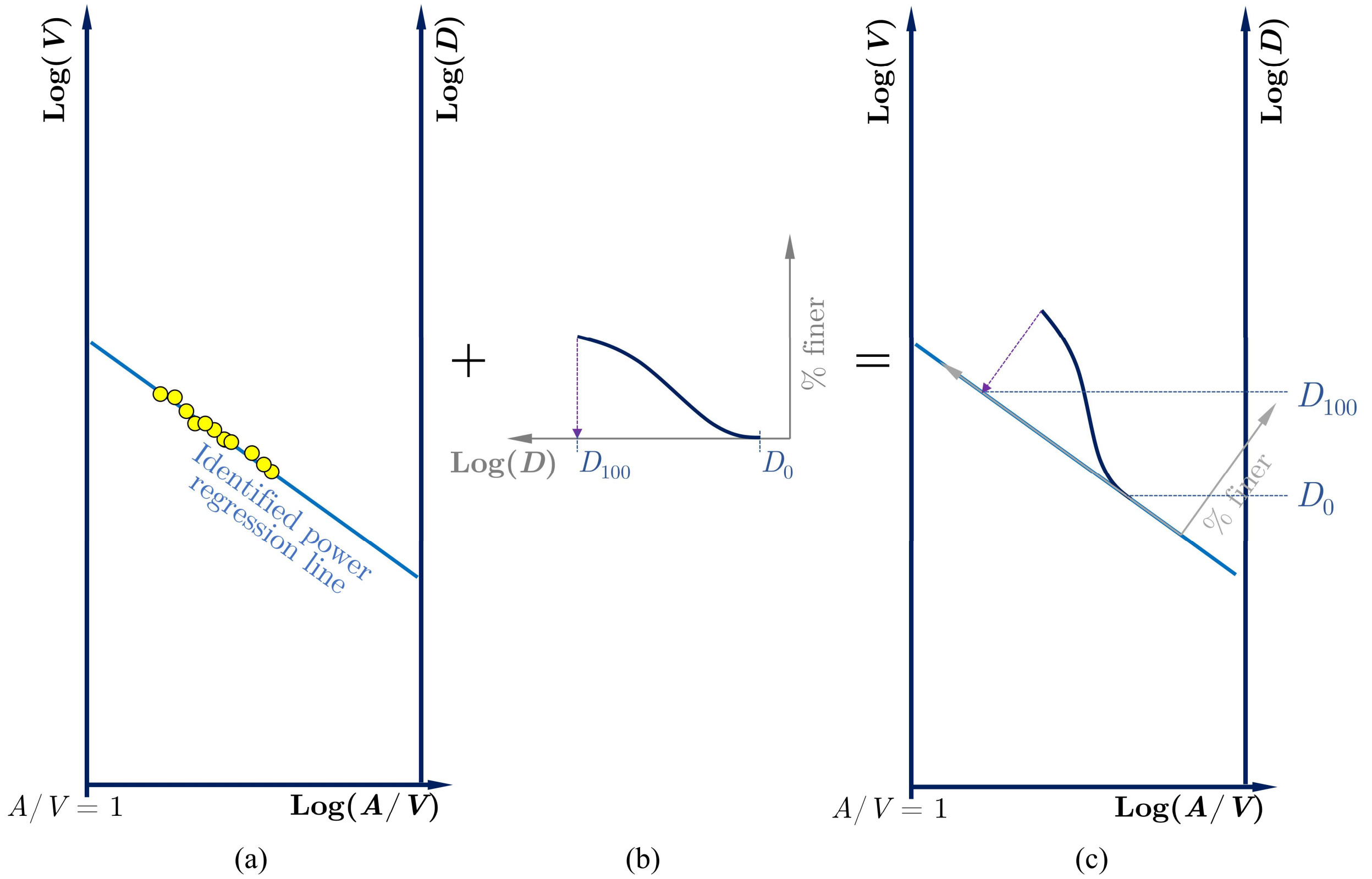

Figure 10. Integration of particle size distribution (PSD) into particle geometry space (PGS): (a) Power regression line identified for a group of particles within PGS; (b) PSD; and (c) Scaling the PSD based on size $D$ and aligning it with the power regression line.

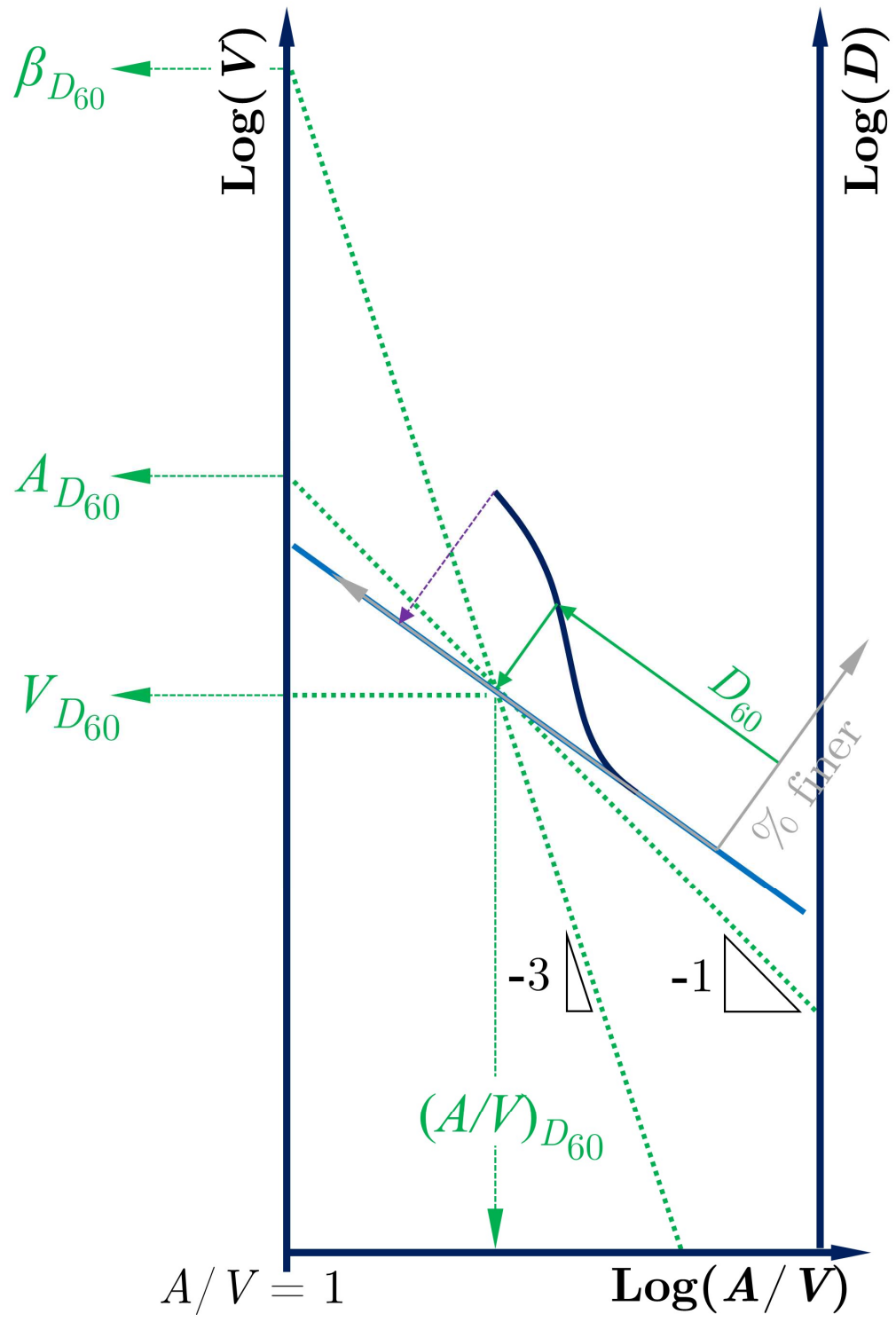

Figure 11. Evaluation of 3D particle geometry attributes corresponding to the PSD, specifically demonstrating the attributes for the particle size $D_{60}$.

# 3 DEMONSTRATION

A set of crushed granite particles, sourced from a quarry near Richmond, Virginia, is used to demonstrate the proposed approach. The PSD of the particles is shown in Figure 12, obtained after sieve analysis. The $D_{10}$, $D_{30}$, $D_{60}$ values are 6.13 mm, 8.88 mm, and 12.19 mm, respectively, and $C_u$ and $C_c$ values are 1.99 and 1.06.

A total of 100 particles of various sizes are randomly selected from the batch for 3D scanning. A Polyga C504 structured light 3D scanner is used, providing a high resolution and accuracy down to 6 microns [24]. Structured light scanning captures 3D models at a 1:1 scale [30], allowing direct measurement of geometric attributes from the digital model without the need for rescaling. The scanning setup is shown in Figure 13. First, a particle is placed on a small putty clump to capture one side at a time (Figure 13a). The particle is then rotated beneath the 3D scanner, positioned directly above, to progressively capture each surface (Figure 13b). About 15 to 20 scans per particle are performed to complete the process. The individual 3D scans are subsequently merged, producing a digital particle model that accurately represents the scanned mineral particle (Figure 13c). A companion software, FlexScan3D, is used for post-processing, including the removal of background artifacts such as the putty used for support. The entire process typically takes about 5 to 10 minutes per particle. From the resulting digital model, the particle's surface area ($A$) and volume ($V$) are measured. The complete dataset of the 100 particles, including the 3D digital particle files, is available as 'Virginia Granite - Group A' in the NSF DesignSafe-CI repository [25].

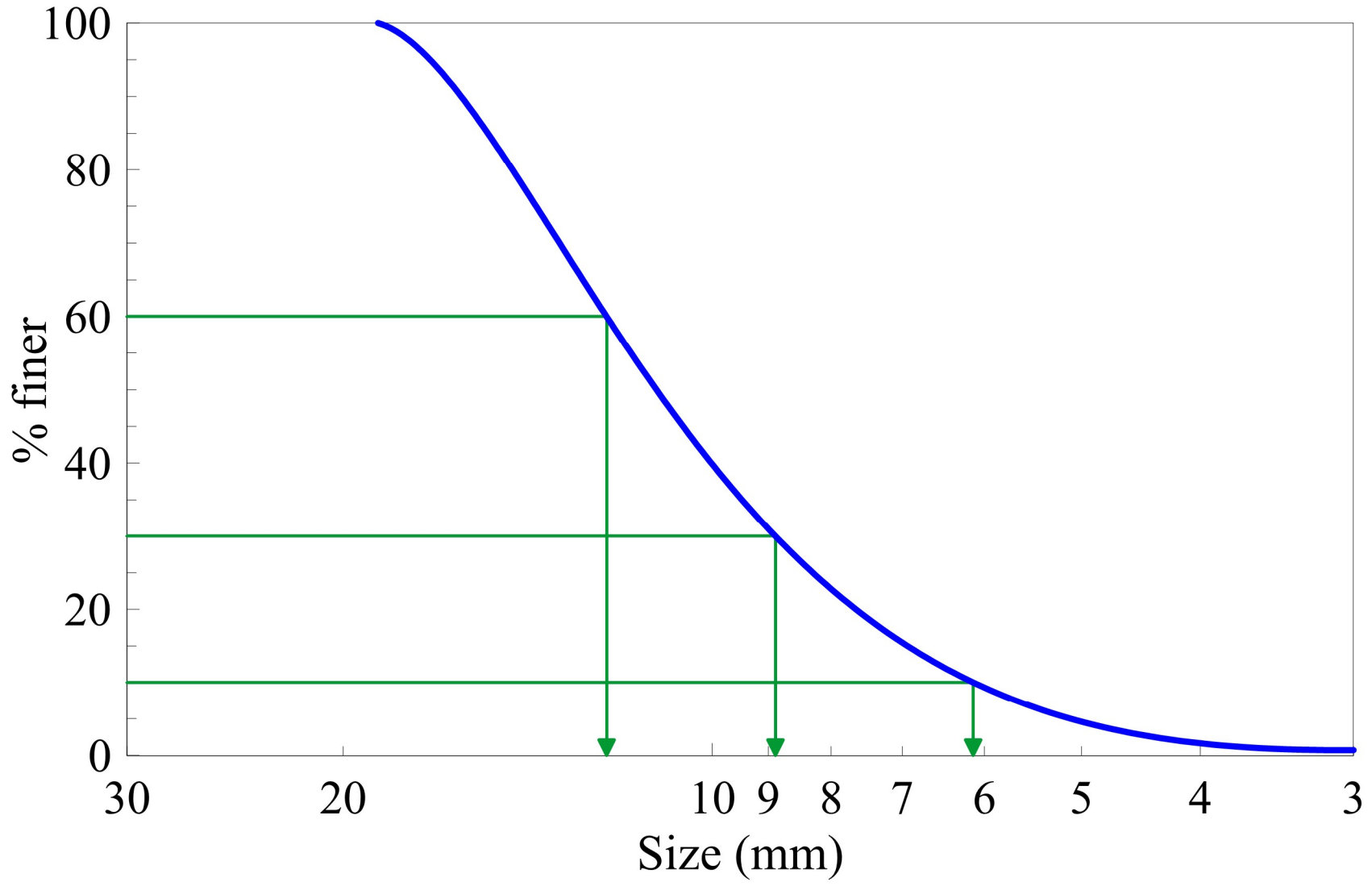

Figure 12. Particle size distribution (PSD) of the studied particles, with $D_{10}$, $D_{30}$, and $D_{60}$ highlighted.

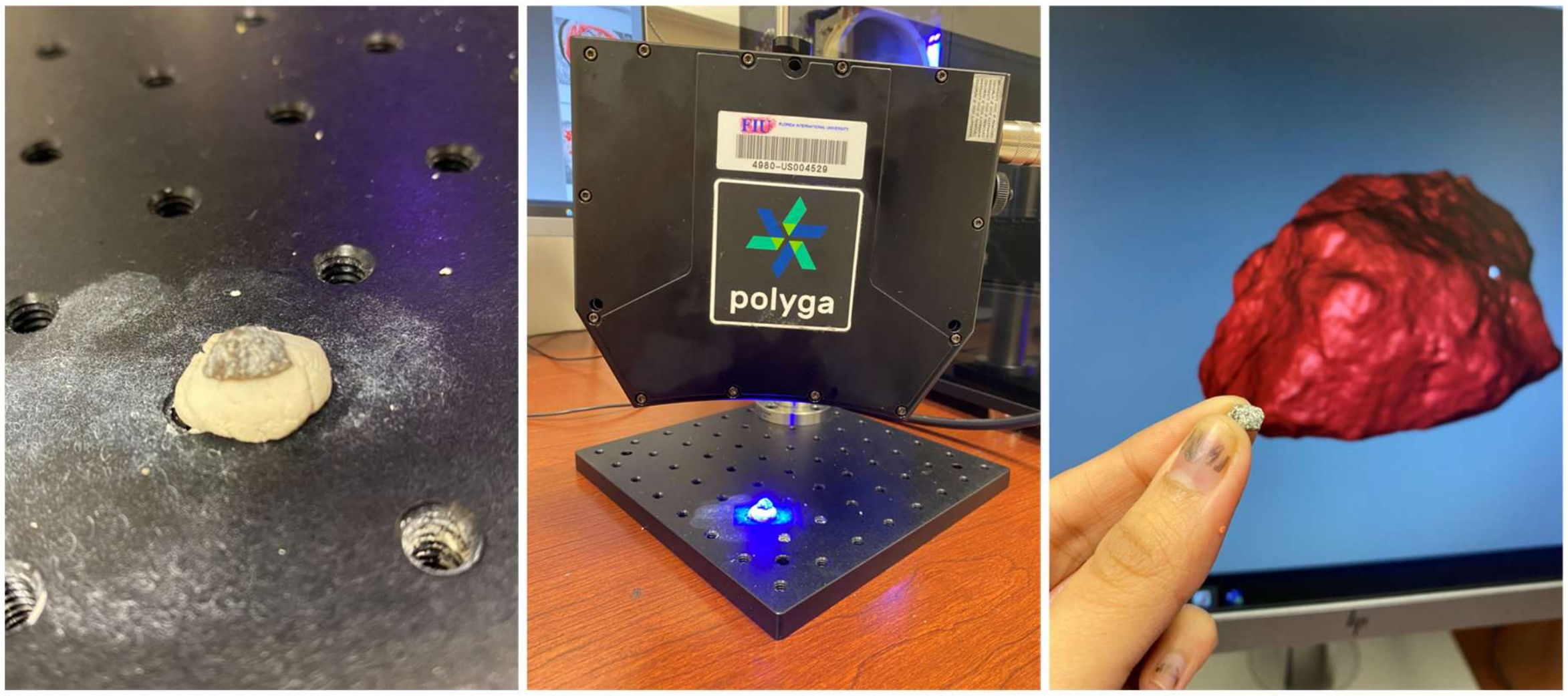

Figure 13. Particle scanning process: (a) Particle positioned on putty for stability; (b) 3D scanning using Polyga C504 structured light 3D scanner; and (c) Resulting 3D digital particle displayed on the screen; The image is sourced from Tripathi et al. [25] and is available under the Open Data Commons Attribution License.

Figure 14 shows the particle geometry data obtained from the sampled particles. The identified power regression line is expressed as $V = (A/V)^{-2.622} \times 274.793$, indicating $\alpha = -2.622$, which corresponds to Case 2 discussed in Section 2.2. The $\beta^*$ value is 299.923, representing the average shape of the particle set. The coefficient of determination ($R^2$) is 0.91. It is important to clarify that we are not suggesting sampling of 100 particles is needed to determine a power regression. Typically, a smaller number of particles is sufficient when the data exhibits strong coherence along the power regression line, as indicated by a high $R^2$ value, provided that the particles share a common geological history.

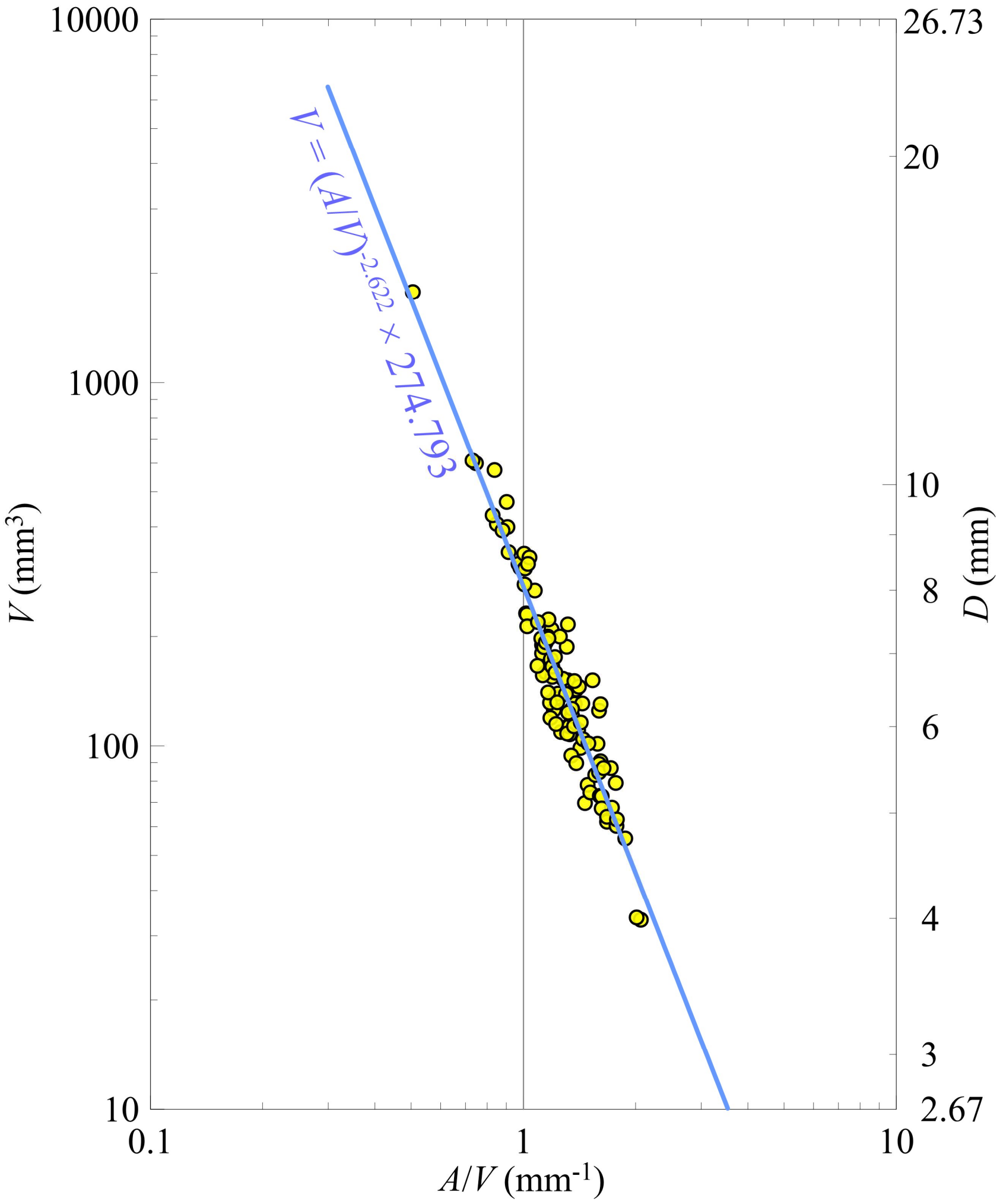

Figure 14. Particle geometries derived from 3D scanning of sample particles, along with the identified power regression line.

Figure 15 demonstrates how the PSD is integrated into the PGS, aligning with the identified power regression line after being scaled using sizes of 20 mm and 3 mm. Particle shape corresponding to a specific size can then be graphically determined. For example, a line with a slope of -3 can be drawn from $D_{60}$ (= 12.19 mm). The point where this line intersects $A/V = 1$ represents its shape $\beta_{D_{60}}$. Similarly, the surface area $A_{D_{60}}$ can be graphically determined from a line with a slope of -1. The volume $V$ for $D_{60}$ can simply be read from the $y$-axis ($V_{D_{60}}$ in the figure), and the $A/V$ value from the $x$-axis.

The values can be computed more accurately using Equations (2) to (5). The volume $V$ for $D_{60}$ can be determined using Equation (2), yielding 948.44 mm$^3$. The $A/V$ value can then be obtained from the power regression function $V = (A/V)^{-2.622} \times 274.793$, resulting in 0.623 mm$^{-1}$. The shape index $\beta_{D_{60}}$ can be calculated using Equation (5) as $(A/V)^3 \times V$, yielding 229.85. Similarly, the surface area $A_{D_{60}}$ can be computed as $(A/V) \times V$, resulting in 591.32 mm$^2$.

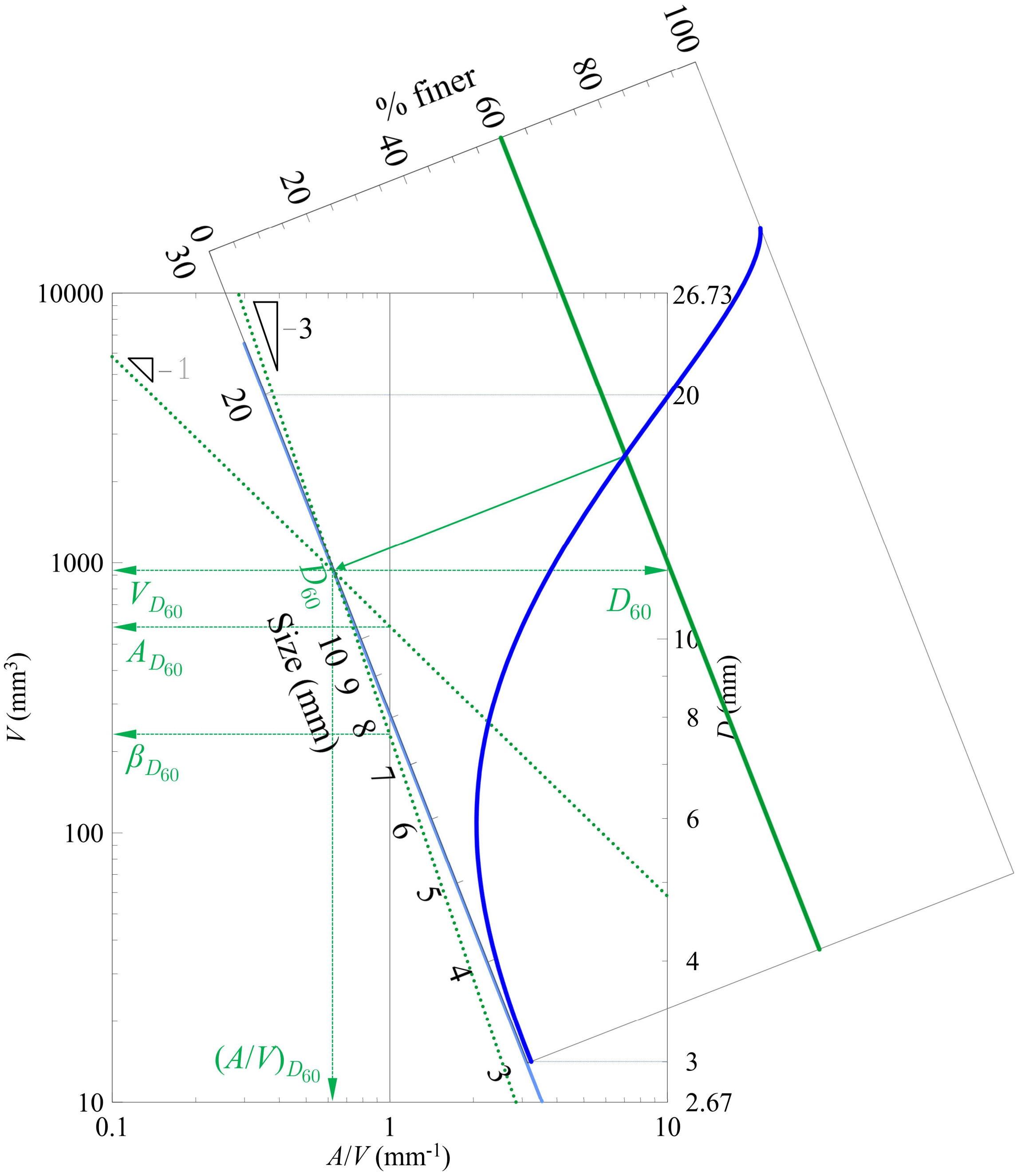

Figure 15. Particle size distribution (PSD) integrated into the particle geometry space (PGS).

## 4 CONCLUDING REMARKS

This study introduces a new approach that integrates particle shape ($\beta$), surface area ($A$), volume ($V$), specific surface ($A/V$), and particle size distribution (PSD) within a unified framework, termed *particle geometry space* (*PGS*). Salient points are summarized as follows:

- PGS is formulated based on the principle that key 3D particle geometry attributes ($\mathscr{g}$) – including volume, surface area, and shape – can be uniformly expressed as a function of specific surface ($A/V$), specifically, $\mathscr{g} = (A/V)^k \times V$. Here, the exponent ($k$) is a set of indices {0,1,3}, each associating $\mathscr{g}$ with a specific 3D geometric attribute.
- This space is logarithmic, with $x$-axis representing log($A/V$), and $y$-axis representing log($V$).
- Within PGS, intercepts at $A/V = 1$ of lines extending from a particle data point with slopes 0, -1, and -3 represent $V$, $A$, and $\beta$, respectively.
- The geometries of particle population can be graphically represented in PGS, capturing both (i) the average shape and (ii) the overall relationship between shape and size, utilizing the power regression line derived from the geometry data.
- PSD can be integrated into PGS by aligning it with an identified power regression line. Subsequently, this facilitates the determination of other 3D particle geometry attributes for any size within PSD.

This approach enables comprehensive characterization of all 3D particle geometry attributes, and their relationships with PSD within a unified framework. It transforms the conventional methods that analyze particle shape and size in isolation, often leading to a significant knowledge gap about the overall impact of particle geometry on granular material behavior.

While this framework is transformative, two minor limitations should be noted. First, the proposed PGS considers shape solely through the true sphericity concept, represented by $\beta$, without accounting for other shape indices such as flatness or elongation ratios. As the first research effort to comprehensively characterize all key 3D particle geometry attributes within a unified framework, this study lays a strong foundation for future enhancement. Building upon this framework, the PGS could be further expanded, for instance, into a three-dimensional space by introducing an additional axis for another shape index, thereby enhancing its analytical depth and applicability. Second, in traditional geotechnical engineering practice, sieve analysis determines particle size based on passage through a mesh, whereas in this study, particle size $D$ is volume-based. While these values are generally expected to be close, discrepancies can arise. A future study is recommended to quantify these differences and assess their implications for the applicability of the proposed method.

We encourage the granular materials research community to explore the potential of this framework in investigating complex phenomena related to particle geometry effects. Expanding research efforts to include mechanical tests – such as triaxial compression test, resilient modulus test, or permeability test – would enable valuable correlations between particle geometry attributes (shape, surface area, specific surface, and PSD) characterized within the PGS and the mechanical behavior of granular materials.

In particular, granular materials in railway tracks and pavement subbases undergo permanent deformation from repeated loading. PGS offers a powerful tool for tracking changes in particle shape and size due to breakage or abrasion, linking these changes to long-term deformation phenomena such as settlement and rutting. By leveraging these insights, engineers can enhance predictive capabilities, optimizing the design and maintenance of granular material systems. Further research in this direction will not only validate this PGS framework but also deepen understanding of particle-scale effects on macroscopic behavior, ultimately improving the performance and durability of granular materials in engineering applications.

## ACKNOWLEDGEMENTS

This work was supported by the US National Science Foundation under the award CMMI #1938431. The authors extend their appreciation to Dr. Eric Koehler at Titan America LLC for providing the granite particles analyzed in this study. The opinions, findings, conclusions, or recommendations expressed in this article are solely those of the authors and do not necessarily reflect the views of the agencies.

## DATA AVAILABILITY

The dataset used for the demonstration is available at https://doi.org/10.17603/ds2-p634-pg95 in the US National Science Foundation's DesignSafe Cyber-Infrastructure (www.designsafe-ci.org) repository under the Open Data Commons Attribution License.

## AUTHOR CONTRIBUTION

Conceptualization: SJL; Methodology: SJL; Formal analysis and investigation: PT, SJL; Writing – original draft preparation: PT, SJL; Writing – Review and Editing: SJL; Funding acquisition: SJL; Resources: SJL; Supervision: SJL.

## COMPETING INTERESTS

The authors have no conflicts of interest to declare that are relevant to the content of this article.

# REFERENCES

[1] Qian Y, Lee SJ, Tutumluer E, Hashash YMA, Ghaboussi J. Role of Initial Particle Arrangement in Ballast Mechanical Behavior. Int J Geomech 2018;18:04017158. https://doi.org/10.1061/(ASCE)GM.1943-5622.0001074.

[2] Qian Y, Lee SJ, Tutumluer E, Hashash YMA, Mishra D, Ghaboussi J. Simulating Ballast Shear Strength from Large-Scale Triaxial Tests. Transp Res Rec J Transp Res Board 2013;2374:126–35. https://doi.org/10.3141/2374-15.

[3] Tutumluer E, Hashash YMA, Ghaboussi J, Qian Y, Lee SJ, Huang H. Discrete Element Modeling of Railroad Ballast Behavior. Urbana, Illinois: 2018.

[4] Qian Y, Boler H, Moaveni M, Tutumluer E, Hashash YMA, Ghaboussi J. Characterizing Ballast Degradation through Los Angeles Abrasion Test and Image Analysis. Transp Res Rec J Transp Res Board 2014;2448:142–51. https://doi.org/10.3141/2448-17.

[5] Tripathi P, Lee SJ, Shin M, Lee CH. A New Paradigm Integrating the Concepts of Particle Abrasion and Breakage. Geo-Congress 2024, Reston, VA: American Society of Civil Engineers; 2024, p. 272–81. https://doi.org/10.1061/9780784485309.028.

[6] Wadell H. Volume, Shape, and Roundness of Quartz Particles. J Geol 1935;43:250–80.

[7] Meng M, Duan X, Shi J, Jiang X, Cheng L, Fan H. Influence of particle gradation and morphology on the deformation and crushing properties of coarse-grained soils under impact loading. Acta Geotech 2023. https://doi.org/10.1007/s11440-023-01989-z.

[8] Dinh BH, Nguyen A-D, Jang S-Y, Kim Y-S. Evaluation of erosion characteristics of soils using the pinhole test. Int J Geo-Engineering 2021;12:16. https://doi.org/10.1186/s40703-021-00145-4.

[9] Zheng J, Hryciw RD. Segmentation of contacting soil particles in images by modified watershed analysis. Comput Geotech 2016;73:142–52. https://doi.org/10.1016/j.compgeo.2015.11.025.

[10] Zhao B, Wang J. 3D quantitative shape analysis on form, roundness, and compactness with μCT. Powder Technol 2016;291:262–75. https://doi.org/10.1016/j.powtec.2015.12.029.

[11] Hryciw RD, Zheng J, Ohm H-S, Li J. Innovations in Optical Geocharacterization. Geo-Congress 2014 Keynote Lect. Geo-Characterization Model. Sustain., 2014, p. 97–116. https://doi.org/10.1061/9780784413289.005.

[12] Yang J, Luo XD. Exploring the relationship between critical state and particle shape for granular materials. J Mech Phys Solids 2015;84:196–213. https://doi.org/10.1016/j.jmps.2015.08.001.

[13] Fernlund JMR. Image analysis method for determining 3-D shape of coarse aggregate. Cem Concr Res 2005;35:1629–37. https://doi.org/10.1016/j.cemconres.2004.11.017.

[14] Garboczi EJ, Bullard JW. 3D analytical mathematical models of random star-shape particles via a combination of X-ray computed microtomography and spherical harmonic analysis. Adv Powder Technol 2017;28:325–39. https://doi.org/10.1016/j.apt.2016.10.014.

[15] Altuhafi FN, Coop MR. Changes to particle characteristics associated with the compression of sands. Géotechnique 2011;61:459–71. https://doi.org/10.1680/geot.9.P.114.

[16] Lee SJ, Shin M, Lee CH, Tripathi P. Phenotypic trait of particle geometries. Granul Matter

2022;24:79. https://doi.org/10.1007/s10035-022-01240-8.

[17] Su YF, Bhattacharya S, Lee SJ, Lee CH, Shin M. A new interpretation of three-dimensional particle geometry: M-A-V-L. Transp Geotech 2020;23:100328. https://doi.org/10.1016/j.trgeo.2020.100328.

[18] Holtz RD, Kovacs WD, Sheahan TC. An Introduction to Geotechnical Engineering. 2nd ed. Pearson; 2010.

[19] Tripathi P, Lee SJ, Lee CH, Shin M. Towards 3D Shape Estimation from 2D Particle Images: A State-of-the-Art Review and Demonstration. KONA Powder Part J 2024:2025017. https://doi.org/10.14356/kona.2025017.

[20] Planinšič G, Vollmer M. The surface-to-volume ratio in thermal physics: from cheese cube physics to animal metabolism. Eur J Phys 2008;29:369–84. https://doi.org/10.1088/0143-0807/29/2/017.

[21] Lee SJ, Lee CH, Shin M, Bhattacharya S, Su YF. Influence of coarse aggregate angularity on the mechanical performance of cement-based materials. Constr Build Mater 2019;204:184–92. https://doi.org/10.1016/j.conbuildmat.2019.01.135.

[22] Bird RB, Stewart WE, Lightfoot EN. Transport Phenomena. Revised 2n. Wiley; 2006.

[23] Hirasaki George J. Lecture Notes on Adsorption. Houston, TX: Rice University; 2004.

[24] Polyga. Polyga Compact C504. PolygaCom 2021. https://www.polyga.com/products/compact-c504/.

[25] Tripathi P, Lee SJ, Shin M, Lee CH. Replication Data: 3D Geometry Characterization of Florida and Virginia Mineral Particles. US Natl Sci Found Des 2023. https://doi.org/10.17603/Ds2-P634-Pg95.

[26] Bhattacharya S, Subedi S, Lee SJ, Pradhananga N. Estimation of 3D Sphericity by Volume Measurement – Application to Coarse Aggregates. Transp Geotech 2020;23:100344. https://doi.org/10.1016/j.trgeo.2020.100344.

[27] Abu-Haifa M, Lee SJ, Zhu C. Phenotypic Trait of Painting Cracks. Stud Conserv 2024:1–16. https://doi.org/10.1080/00393630.2024.2420574.

[28] Hardin BO. Crushing of Soil Particles. J Geotech Eng 1985;111:1177–92. https://doi.org/10.1061/(ASCE)0733-9410(1985)111:10(1177).

[29] Einav I. Breakage mechanics—Part I: Theory. J Mech Phys Solids 2007;55:1274–97. https://doi.org/10.1016/j.jmps.2006.11.003.

[30] Lansdown H. Digital Modelmaking: Laser Cutting, 3D Printing and Reverse Engineering. Crowood Press; 2019.